\documentclass[aps,pre,showpacs,showkeys,preprint]{revtex4}

\usepackage{graphicx}
\usepackage{latexsym}

\begin{document}

\title{Generation of scale-free networks using a simple preferential rewiring dynamics}

% Authors information
\author{Suhan Ree}
\email{suhan@kongju.ac.kr}
\affiliation{Department of Industrial Information,
Kongju National University, Yesan-Up, Yesan-Gun, 
Chungnam, 340-702, South Korea}

\date{\today}

\begin{abstract}
We propose a simple dynamical model that generates networks with power-law degree distributions with the exponent 2
through rewiring only.
At each time step, two nodes, $i$ and $j$, are randomly selected, and one incoming link to $i$
is redirected to $j$ with the rewiring probability $R$, 
determined only by degrees of two nodes, $k_i$ and $k_j$,
while giving preference to high-degree nodes.
To take the structure of networks into account,
we also consider what types of networks are of interest, whether links are directed or not, 
and how we choose a rewiring link out of all incoming links to $i$,
as a result, specifying 24 different cases of the model.
We then observe numerically that networks will evolve to steady states with power-law degree distributions
when parameters of the model satisfy certain conditions.
\end{abstract}

% 89.75.Da : Systems obeying scaling laws
% 89.75.Hc : Networks and genealogical trees


\maketitle

%\setcounter{totalnumber}{1}

% Body of the paper starts here
% ==============================<Introduction>==================================================
%\section{Introduction}
Complex systems have been studied in various research fields as a general framework
to describe systems that usually consist of interacting elements,
and, in most cases, these systems as a whole exhibit emerging behavior.
We can represent complex systems as networks with nodes (or vertices) and links (or edges),
where nodes represent elements of a given system, and links represent interactions (physical or conceptual) 
between two elements.
Recently, structural and dynamical properties of these networks and their underlying mechanisms
have become the focal point of many researches \cite{strogatz,albert3,dorogovtsev2,newman2}.
One well-known property of many natural and artificial networks alike
is the power-law behavior of degree distributions $P(k)$ (the degree $k$ of a node is the number of connected links),
and such networks are called {\em scale-free} networks; in other words, 
$P(k)\propto k^{-\gamma}$ where $\gamma$ is the exponent, usually in the range of $2\le\gamma\le3$.
Examples are citations of scientific papers \cite{redner}, links in web pages \cite{albert}, the Internet \cite{faloutsos}, 
metabolic networks \cite{jeong}, and brain functional networks \cite{eguiluz}, and so on.

Many generating mechanisms have been proposed so far to explain this phenomenon.
One popular approach is using stochastic dynamical processes, in which networks evolve with time and approach
scale-free states, ensembles of network structures sharing specific statistical properties.
If we classify these mechanisms into two classes, one is growing mechanisms, where
both the number of nodes, $N$, and the number of links, $E$, increase with time,
and the other is non-growing mechanisms, where $N$ and $E$ are either constant or varying but limited
to finite values.
One of the most studied growing mechanisms is 
the preferential-attachment (PA) mechanism \cite{barabasi}, 
where a node and  a fixed number of links connected to this node are added at each time step,
while other ends of these links are chosen from existing nodes with the probability proportional to their degrees.
There are many generalizations and extensions of PA mechanism, usually with additional features: for example,
rewiring \cite{albert2}, attractiveness of nodes \cite{dorogovtsev}, linking between existing nodes \cite{krapivsky},
age and deactivation \cite{klemm}, local knowledge \cite{gomez}, and link weights \cite{barrat}.
While PA mechanism is valid to many scale-free growing networks, it is not suitable
for non-growing networks,
and there also are numerous non-growing mechanisms that use some additional features or
information: for example,
transitive linking and aging \cite{davidsen},
fitness of nodes \cite{caldarelli},
adding links using nonlinear PA \cite{mukherjee},
local optimization using memories of nodes \cite{rosval},
optimization of Hamiltonians \cite{baiesi},
and merging and regeneration of nodes \cite{kim}.
Models mentioned so far share the property of power-law degree distributions, but
they also attempt to capture different aspects of various types of real networks, which we don't describe details here.

In this paper, we propose a non-growing mechanism (both $N$ and $E$ are constant) with
a stochastic preferential-rewirinig process, where 
only degree information of interacting nodes is used for the dynamics.
In other words, nodes and links have no variables that represent memory, attractiveness,
or weight, and the global structural knowledge of the network is not needed,
thereby making the dynamics {\em simple} and {\em local}.
% ==============================<Model>===================================================
%\section{Model}
Our network model is an application of the general model introduced in Ref.\ \cite{ree},
which we first describe briefly.

The general model deals with $N$-element systems, where each element $i$ is represented
by a quantity $k_i$, a non-negative integer. 
At each time step, two elements, $i$ and $j$, are randomly 
chosen out of $N$ elements and one unit of quantity is given from $i$ to $j$ with the
exchange probability $R$, determined only by $k_i$ and $k_j$.
When the exchange occurs, $k_i$ and $k_j$ are changed to $k_i-1$ and $k_j+1$
(the average quantity, $\alpha\equiv\sum_{i=1}^N k_i/N=\langle k \rangle$, 
is constant). 
The exchange probability is defined using a parameter $\beta$ as below,
\begin{equation}
	R = \left\{
		\begin{array}{lc}
			1  &  (0 < k_i \leq k_j) \\
			\beta   & (k_i > k_j) \\
			0  & (k_i = 0) \\
		\end{array}\right.,
\label{r}
\end{equation}
where $0\le \beta \le 1$.
This is a generalized zero-range process (ZRP) in a fully connected geometry \cite{evans},
even though $R$ as a function of $k_i$ and $k_j$ in Eq.\ (\ref{r}) is rather unique.
It was found analytically and numerically that the distribution of $k_i$'s
exhibits the power law when $\alpha$ and $\beta$ satisfy the condition in the limit of large $N$,
\begin{equation}
	\alpha_c(\beta)=\frac{\beta}{1-\beta}\ln\!\left[\epsilon^{-1}\beta^{\frac{1}{1-\beta}}\right]<\alpha,
\label{alphac}
\end{equation}
where $\epsilon$ is a small positive constant, estimated to be about $10^{-3}$.
This equation defines a power-law regime inside $(\alpha,\beta)$-space;
i.e., to have a power-law distribution when $\beta$ is given,
$\alpha$ should be greater than the critical value, $\alpha_c(\beta)$
[or when $\alpha$ is given, $\beta$ should be less than the critical value, $\beta_c(\alpha)$, which is
the inverse of $\alpha=\alpha_c(\beta)$].
For example, from $\alpha_c(0.5)\simeq6$,
we deduce that $\beta$ should be less than 0.5
to have a power-law distribution when $\alpha=6$. 
When Eq.\ (\ref{alphac}) is satisfied, 
the stationary probability distribution, $P(k)$, and the stationary cumulative distribution, $P(\ge\!k)$, have
been found analytically as below,
\begin{eqnarray}
P(k)  &=& \frac{\beta }{{1 - \beta }}\frac{1}{{[k + 1/(1 - \beta )][k + \beta /(1 - \beta )]}},\nonumber\\
P(\ge\!k)  &=& \frac{\beta }{{1 - \beta }}\frac{1}{{k + \beta /(1 - \beta )}},
\label{PkECkE}
\end{eqnarray}
which is the power-law distribution with $\gamma=2$ when $k(1-\beta)\gg 1$.

\begin{figure}
\includegraphics[scale=0.5]{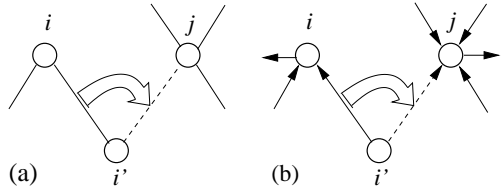}
\caption{\label{model}
A diagram describing the rewiring dynamics. 
Links of the given network can be either (a) undirected, or (b) directed.
For both cases, two nodes, $i$ and $j$, are randomly chosen ($k_i=2$ and $k_j=3$), 
and the node $i'$ is one of the neighbors of $i$.
(For networks with directed links, $k_i$ represents the in-degree of the network,
and $i'$ is one of the neighbors with an outgoing link to $i$.)
Rewiring from $(i',i)$ to $(i',j)$ can occur with the rewiring probability $R$.
}
\end{figure}

A network with undirected links ($\alpha=2E/N$) is an example of such $N$-element systems, where $k_i$ represents the degree
of a node $i$ and $R$ is the {\em rewiring} probability
(for networks with directed links, $k_i$ represents the in-degree, the number of incoming links, of a node and $\alpha=E/N$).
In Fig.\ \ref{model}, the rewiring dynamics is described using a schematic diagram.
%===========Figure 1============================
%===================================================
A rewiring process involves three nodes: $i$ and $j$ are randomly chosen, and
$i'$, a pivot node, is a neighbor of $i$, which means a node directly connected to $i$.
When $k_i$ and $k_j$ are given, rewiring from $(i',i)$ to $(i',j)$ can occur
with the rewiring probability $R$ in Eq.\ (\ref{r}).
Due to the presence of the pivot node and different network types, our network model should be more than just a ZRP
if we consider the link structure of networks,
and we introduce three additional distinctions:
how to choose $i'$ among neighbors of $i$ ({\em pivot type}), whether links are directed or not ({\em link type}), and 
whether loops and/or multiple links are allowed or not ({\em network type}).
(When two end nodes of a link are the same, we call it a loop, and when there are more than one link
between two nodes, we call them multiple links).

Here we introduce three pivot types: (i) the node with the smallest
degree (`$S$'), (ii) the randomly chosen node (`$R$'), (iii) the node with the largest degree (`$L$').
There are two link types: (i) undirected (`$U$'), (ii) directed (`$D$').
And there can be four types of networks: (i) both loops and multiple links are allowed (`1'), (ii) only loops
are allowed (`2'), (iii) only multiple links are allowed (`3'), (iv) neither a loop nor multiple links are allowed (`4').
Then there are 24 different cases of the model.
For convenience, we will use a three-letter notation from here on, 
where the first letter represents the pivot type, 
the second letter represents the link type, and the last letter represents the network type.
For example, $SD4$
represents networks with directed links, which allow neither a loop nor multiple links,
when $i'$ is the neighbor of $i$ with the smallest degree.
In addition, three more details should be mentioned:
(i) while choosing the pivot node, if there are more than one node with the smallest or largest degree,
one node is chosen randomly out of them,
(ii) when the link type is $D$ (directed), out-degrees of neighbors are used when choosing the pivot node,
(iii) when the network type is not 1, a rewiring attempt can be aborted if it results in a loop or multiple links.

\begin{figure}
\includegraphics[scale=0.5]{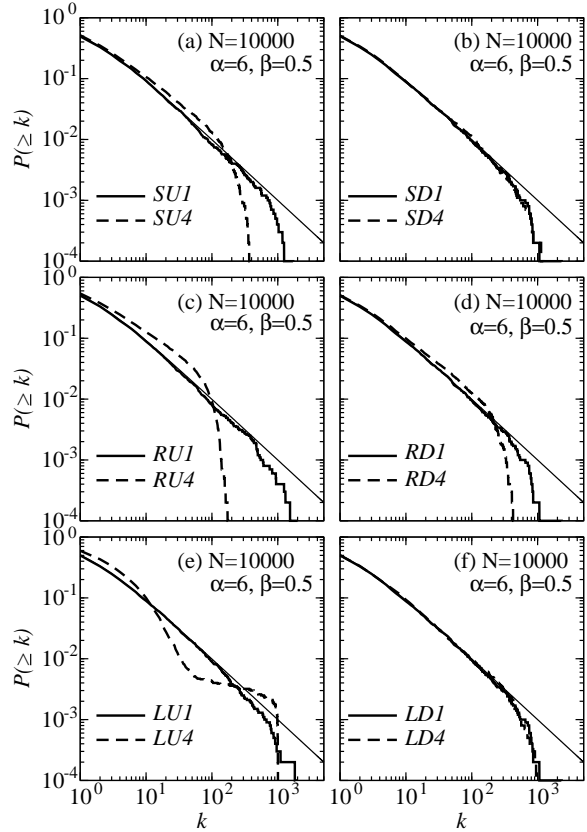}
\caption{\label{alltypes}
Stationary cumulative degree distributions,
$P(\ge\!k)$, of 12 different cases,
when $\alpha=6$, $\beta=0.5$, and $N=10000$: 
(a) $SU1$ and $SU4$, (b) $SD1$ and $SD4$, (c) $RU1$ and $RU4$,
(d) $RD1$ and $RD4$, (e) $LU1$ and $LU4$, (f) $LD1$ and $LD4$.
%When the network type is 1, the stationary degree distributions
%are the same for all cases.
%When the network type is 4, the stationary degree distributions depend on 
%the link type and the pivot type.
The thin solid line represents the analytic solution, $P(\ge\!k)=1/(1+k)$, 
of the general model when $\beta=0.5$ as $N\rightarrow\infty$.
}
\end{figure}

% ====================<Numerical Results>========================================
%\section{Numerical Results}
In Fig.\ \ref{alltypes}, we observe stationary degree distributions, $P(\ge\!k)$, of 12 different cases
when $\alpha=6$, $\beta=0.5$, and $N=10000$.
%================Figure 2=======================
%==================================================
If we look at the cases when both loops and multiple links are allowed (network type 1),
degree distributions are the same as those from the general model, because
degree distributions don't depend on the pivot type and the link type
even though other structural properties of the network do.
These cases can be regarded as ZRPs when we are only interested in degree distributions.
On the other hand, when neither loops nor multiple links are allowed (network type 4),
the degree distributions are different from the general model because rewiring attempts can be aborted
depending on the current link structure of the network;
as a result, cutoffs  are usually lower.
When the pivot type is $S$, the distribution is closer to that of the general model even though
cutoffs are still somewhat lower,
because the probability of aborting a rewiring attempt is lower when the pivot node has 
the smallest degree [see Figs.\ \ref{alltypes}(a) and (b)], and
that is especially true for cases with undirected links.
When the network type is 2 or 3
(not shown here),
results are almost the same as those of the network type 4 or 1, 
respectively.
This shows us that the presence of loops does not have much impact on stationary distributions.

\begin{figure}
\includegraphics[scale=0.5]{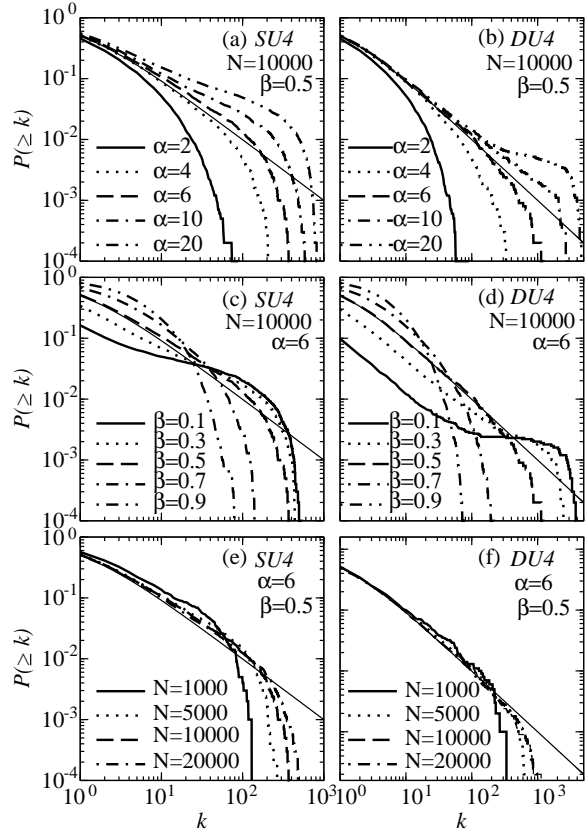}
\caption{\label{type4}
The stationary cumulative degree distributions, $P(\ge\!k)$, of networks, $SU4$ and $SD4$, when
one of three parameters of the model, $\alpha$, $\beta$ and $N$, is varied.
(a), (b) $\alpha$ is varied;
(c), (d) $\beta$ is varied; and
(e), (f) $N$ is varied.
The thin solid line represents the analytic solution, $P(\ge\!k)=1/(1+k)$, 
of the general model when $\beta=0.5$ as $N\rightarrow\infty$.
%When $\alpha\simeq\alpha_c(\beta)$,
%the degree distributions get close to the theoretical curve, and as $N$ grows the 
%cutoff increases.
}
\end{figure}

In Fig.\ \ref{type4}, we observe the behavior of the distributions of the $SU4$ and $SD4$ networks
more closely.
%================Figure 3=======================
%==================================================
We vary one parameter while the other two parameters are fixed.
In Figs.\ \ref{type4}(a) and (b), $\alpha$ is varied when $N$ and $\beta$ are fixed,
while in Figs.\ \ref{type4}(c) and (d), $\beta$ is varied when $N$ and $\alpha$ are fixed.
We observe that the stationary degree distributions are close to the
theoretical curve when $\alpha$ is close to $\alpha_c(\beta)$ [or $\beta$ is close to $\beta_c(\alpha)$] unlike
the general model, 
where stationary distributions were found to follow the power law
when $\alpha>\alpha_c(\beta)$ [or $\beta<\beta_c(\alpha)$].
In Figs.\ \ref{type4}(e) and (f), $N$ is varied when $\alpha$ and $\beta$ are fixed.
We observe that cutoffs increase as $N$ increases, because the sparser the network gets, 
the smaller the chance of aborting the rewiring attempt becomes.

\begin{figure}
\includegraphics[scale=0.5]{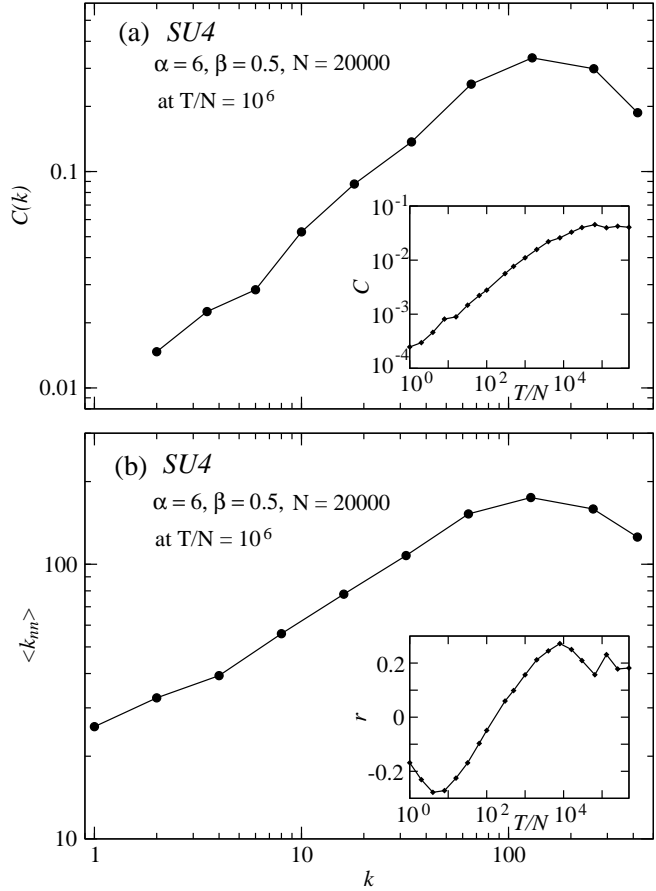}
\caption{\label{correlation}
Structural properties of stationary $SU4$ networks when $\alpha=6$, $\beta=0.5$, and $N=20000$.
%When an Erd\"{o}s-R\'{e}nyi random network is used as the initial network, the steady state is reached
%at $T/N\sim10^5$.
Data points ($\bullet$) are obtained from averaging values of individual nodes in log-sized bins in both cases.
(a) The averaged clustering coefficient, $C(k)$, versus $k$ in a log-log plot at $T/N=10^6$.
Inset: The clustering coefficient, $C$, versus $T/N$ in a log-log plot.
(b) The averaged degree of nearest neighbors of nodes with degree $k$, 
$\langle k_{nn}\rangle$, versus $k$ in a log-log plot at $T/N=10^6$.
It shows that the stationary network is assortatively mixed.
Inset: The Pearson correlation coefficient of the degrees, $r$, versus $T/N$ in a semi-log plot.
}
\end{figure}

In Fig.\ \ref{correlation}, we observe other structural properties of the $SU4$ network
of the steady state when $\alpha=6$, $\beta=0.5$, and $N=20000$.
%================Figure 4=======================
%==================================================
There are several well-known statistical measures \cite{albert3,dorogovtsev2,newman2} such as the clustering coefficient, $C$ \cite{cc}, 
the average shortest path length, $L$, and the Pearson correlation coefficient of degrees, $r$ \cite{newman}. 
Here an initial network is the Erd\"{o}s-R\'{e}nyi (ER) random network, where $C=0.0003$, $L=5.73$, and $r=0.00008$.
When the stationary distribution is reached at $T/N\sim 10^5$, we find that 
$C=0.04(0.002)$, $L=4.53(0.03)$, and $r=0.2(0.02)$,
showing the {\em small-world} behavior \cite{watts}.
These networks have more than one component: for ER networks, the size of the largest component is 19958 and $P(0)=0.002$,
and they become 10199(53) and 0.467(0.003) respectively in the steady state.
In Fig.\ \ref{correlation}(a), the average clustering coefficient of nodes with degree $k$, $C(k)$,
is found, and we can observe that $C(k)$ increases with $k$.
And the inset shows how $C$ increases with time until $T/N\sim10^5$.
In Fig.\ \ref{correlation}(b), the averaged degree of nearest neighbors of nodes with degree $k$, $\langle k_{nn}\rangle$,
is found, and we can observe that $\langle k_{nn}\rangle$ increases with $k$ also.
The inset shows how $r$ changes with time, and both results imply that this network is assortatively mixed.
In both cases, the increasing behavior stops at $k\sim200$, and this is probably due to the finite-size effect
[the distribution deviates from the theoretical curve at $k\sim200$ in Fig.\ \ref{type4}(e)].

% ===========================<Discussions>=============================================
%\section{Discussion}
In conclusion, we introduced the rewiring model that generates scale-free networks with $\gamma=2$.
We showed numerical results for 12 different cases of the model, and observed that, when the network can have
loops and multiple links, stationary degree distributions for all cases, including those from
the general model, follow the same power law if the 
condition $\alpha>\alpha_c(\beta)$ is satisfied.
While there exist networks with multiple links (for example, citation network),
networks without a loop and multiple links have been of interest in most studies.
We also observed that, when the network can have neither a loop nor multiple links, scale-free networks
are generated if the condition $\alpha\simeq\alpha_c(\beta)$ is satisfied.
This model was not conceived to describe a certain type of real networks, but 
it was from an effort to find a simple rewiring model generating scale-free networks,
with intention of using local degree information only and incorporating the rich-get-richer phenomenon.
As one of the simplest rewiring models for scale-free networks,
this model can be extended to meet specific needs of some real non-growing networks.

\begin{acknowledgments}
This work was supported by grant No.\ R05-2002-000799-0 from the Basic 
Research Program of the Korea Science \& Engineering Foundation.
\end{acknowledgments}

\bibliography{paper}

\begin{thebibliography}{27}
\bibitem[{cc()}]{cc}
\bibinfo{note}{The definition, $C=\sum_iC_i/N$, where $C_i$ is the ratio of the
  number of links between neighbors of $i$ to $k_i(k_i-1)/2$, is used.}
\end{thebibliography}

\end{document}